\documentclass{article}

\usepackage[final]{neurips_genbio}

\usepackage[utf8]{inputenc} 
\usepackage[T1]{fontenc}    
\usepackage{hyperref}       
\usepackage{url}            
\usepackage{booktabs}       
\usepackage{amsfonts}       
\usepackage{nicefrac}       
\usepackage{microtype}      
\usepackage{xcolor}         
\usepackage{graphicx}
\usepackage{svg}
\usepackage{siunitx}
\usepackage{enumitem}
\usepackage{multirow}

\usepackage{soul, color}
\newcommand{\Note}[1]{}

\setcitestyle{numbers,open={[},close={]},citesep={,}}

\title{Evaluating Zero-Shot Scoring for \textit{In Vitro} Antibody Binding Prediction with Experimental Validation}

\author{
  Divya Nori\thanks{Work done during internship at Absci.}\\
  Massachusetts Institute of Technology\\
    Department of Electrical Engineering and Computer Science\\
  77 Massachusetts Ave, Cambridge, MA 02139 \\
  \texttt{divnor80@mit.edu} \\
  \And
  Simon V. Mathis\footnotemark[1] \\
  University of Cambridge \\
  Department of Computer Science and Technology\\
\texttt{simon.mathis@cl.cam.ac.uk} \\
  \And
  Amir Shanehsazzadeh \\
  Absci Corporation\\
\texttt{ashanehsazzadeh@absci.com} \\
}

\begin{document}

\maketitle

\begin{abstract}
  The success of therapeutic antibodies relies on their ability to selectively bind antigens. AI-based antibody design protocols have shown promise in generating epitope-specific designs. Many of these protocols use an \emph{inverse folding} step to generate diverse sequences given a backbone structure. Due to prohibitive screening costs, it is key to identify candidate sequences likely to bind \textit{in vitro}. Here, we compare the efficacy of $8$ common scoring paradigms based on open-source models to classify antibody designs as binders or non-binders. We evaluate these approaches on a novel surface plasmon resonance (SPR) dataset, spanning $5$ antigens. Our results show that existing methods struggle to detect binders, and performance is highly variable across antigens. We find that metrics computed on flexibly docked antibody-antigen complexes are more robust, and ensembles scores are more consistent than individual metrics. We provide experimental insight to analyze current scoring techniques, highlighting that the development of robust, zero-shot filters is an important research gap.

\end{abstract}



\section{Introduction}

Antibodies are a prominent class of therapeutic agents, primarily because they can selectively bind to a wide array of disease-causing target antigens. This binding specificity is governed by interactions between the antibody's paratope region and the antigen's epitope region \cite{peng2014origins}. Epitope specificity is primarily facilitated by the structure of hypervariable loops in the paratope called complimentarity determining regions (CDRs).

Computationally designing the CDRs such that they bind to specific epitopes with high affinity has become a problem of great interest. Researchers often start with a known or predicted antibody backbone structure, and the task is to design sequences that assume this backbone conformation \cite{hsu2022learning}. Recently, generative machine learning methods have proven effective for this \textit{inverse folding} problem on protein design~\cite{Dauparas2022}. Given the hypervariability of CDRs, there are many sequences that can fold into the same 3D backbone structure. Filtering the diverse outputs of an inverse folding model \textit{in silico} to select for sequences with greater chance of \textit{in vitro} success remains an important challenge, since experimental resources are often limited.

In the protein space, many \textit{in silico} metrics are commonly used to evaluate the quality of designs \cite{johnson2023computational}, \cite{wu2022protein}. Here, we investigate how these metrics perform for antibodies. We benchmark several scoring paradigms on their ability to filter antibody designs by predicted success in a surface plasmon resonance (SPR) assay, an industry standard assay used to measure binding affinity.

\section{Background}

\begin{figure}[h]
  \centering
  \includegraphics[width=\textwidth]{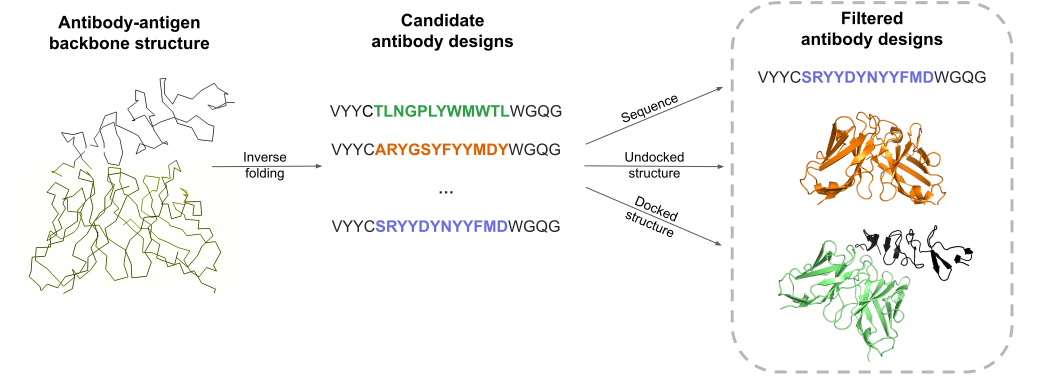}
  \caption{Antibody-inverse folding model with antibody-antigen complex used as input predicts a set of candidate CDR sequences. These sequences are then filtered at the sequence, undocked structure, or docked structure level, which is the focus of this study. The HCDR3 sequences are from~\cite{shanehsazzadeh2023unlocking}.}
  \label{Figure1}
\end{figure}

\paragraph{Problem set-up} As shown in Figure \ref{Figure1}, inverse folding models design candidate sequences given a desired backbone structure. A subset of these sequences will fold into a 3D structure \textit{in vitro}, and a smaller subset then assume paratope residue conformations that interact strongly with the antigen \cite{peng2014origins}.

Here, we assess whether filtering at the sequence level, undocked structure level, or docked structure level is effective. Particularly, we compare $8$ different scoring paradigms on the binder vs non-binder classification task. The candidate scoring paradigms we evaluate are all zero-shot methods -- meaning they are not trained on any binding affinity data which is often available in only limited quantities.

\paragraph{Related work} In general protein binder design, \textit{in silico} metrics have been used to filter libraries and improve binder success rates. Most recently, it was shown that monomer metrics are less effective in predicting binders than complex metrics \cite{bennett2023improving}. On antibodies, prior work assesses Rosetta's predicted free energy of binding ($\Delta\Delta G$) as a binder scoring function and achieves an AUROC of $0.55$ on a library designed for HER2 \cite{mason2019deep}. Recently, a diffusion model was trained to assess the quality of antibody designs by generating and scoring docked poses \cite{Peng2023.09.25.559190}. Language models have also been used to perform efficient evolution on antibodies, indicating that their likelihoods can be used to select predicted binders \cite{hie2023efficient}. We assess whether such approaches are effective, given novel wet-lab insight.

\section{Methods}

\paragraph{Data} To generate the antibody libraries scored in this work, an antibody-specific inverse folding model was given an antibody-antigen complex backbone structure, antibody framework sequences, and antigen sequence as input. The model was then used to design several candidate antibody sequences, specifically all three HCDRs of a reference antibody to the antigen of interest. Libraries were designed for $5$ different antigens, and wet-lab data was generated by SPR. A future publication will describe additional details on data generation. See Table \ref{table:1} for details on library size and \% binders.

\paragraph{Candidate scoring functions} Table \ref{table:2} describes the $8$ candidate scoring functions we evaluate on the binder classification task. We hypothesize that antibody candidate sequences that contain common or predictable motifs are more likely to be binders. Therefore, we benchmark pseudo-perplexity predicted by the ESM-2 650M parameter model as a sequence scoring metric \cite{lin2022language}. Pseudo-perplexity was computed for each sequence, scaled by min-max normalization, flipped to range [-1,0], and translated to range [0,1]. All scoring paradigms going forward are normalized in a similar manner. Subsequently, we hypothesize that antibody candidates with biologically plausible structural qualities are more likely to be successful binders. As an undocked structure scoring metric, we benchmark the residue-level confidence scores given by ABodyBuilder2 (ABB2), an antibody-specific folding model \cite{abanades2023immunebuilder}.

Other discriminative scoring functions used in-practice are self-consistency metrics, such as similarity between the folded antibody candidate and the input backbone structure \cite{wu2022protein}. Here, we benchmark root mean square deviation (RMSD) as a similarity measure, under three Kabsch alignments \cite{kabsch1976solution}.

Given that flexible interactions between antibody and antigen is critical for binding, we benchmark metrics on flexibly docked complexes. To generate these complexes, we apply a previously proposed architecture, dyMEAN~\cite{kong2023end}, and make several modifications to the training procedure which are described in the \hyperref[dymean]{Appendix}. We evaluate three scoring paradigms on the predicted complexes, selected because they capture established heuristics on antibody-antigen interactions. The scores we evaluate measure interface proximity, contact proximity, and interface interactions (details in \hyperref[docked-scoring]{Appendix}).

\begin{table}[h]
\begin{center}
\renewcommand{\arraystretch}{1.3} 
\newcolumntype{P}[1]{>{\centering\arraybackslash}p{#1}} 
\begin{tabular}{ P{3cm} P{4.7cm} P{4.7cm} } 
& \textbf{Model/Method} & \textbf{Metric} \\
\cmidrule(lr){1-3}
\textbf{Sequence Only} & ESM & Pseudo-Perplexity \\
\cmidrule(lr){1-3}
\multirow{4}{*}{\textbf{Undocked Structure}} & 
ABodyBuilder2 & Residue-Level Model Confidence \\ 
& Antibody-Aligned Antibody & RMSD \\ 
& All HCDR-Aligned Antibody & RMSD \\ 
& HCDR3-Aligned Antibody & RMSD \\ 
\cmidrule(lr){1-3}
\multirow{3}{*}{\textbf{Docked Complex}} & Modified dyMEAN & Interface Proximity \\ 
& Modified dyMEAN & Contact Proximity \\ 
& Modified dyMEAN & Interface Interactions \\ 
\end{tabular}
\vspace{0.8em}
\caption{Evaluation of $8$ candidate scoring functions on the binder classification task across $3$ classes of metrics: sequence only, undocked structure, and docked complex.}
\label{table:2}
\end{center}
\end{table}


\vspace{-0.3in}

\section{Results}

\begin{figure}[!ht]
  \centering
  \includegraphics[width=0.9\textwidth]{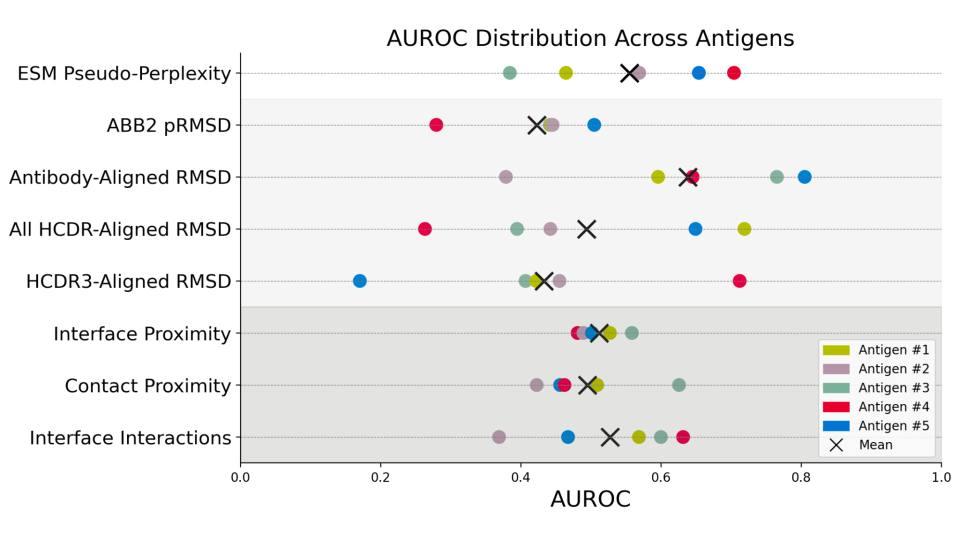}
  \caption{Distribution of AUROC values for various scoring functions across $5$ antigens.}
  \label{Figure2}
\end{figure}

We assess the performance of each scoring paradigm in the classification setting, evaluating across multiple antigens. Figure \ref{Figure2} shows each scoring function's distribution of AUROC values. We observe that antibody-aligned RMSD performs the best (mean of $0.65$ and standard deviation of $0.15$), indicating that undocked structure filtering may be effective. However, all three of the RMSD-based undocked structure metrics display high variance in performance. Interface proximity is the most consistent metric (mean $0.51$ and standard deviation $0.028$), followed by contact proximity and interface interactions. This reveals that docked complex metrics exhibit less variability. 

After evaluating the performance of each metric individually, we assess whether we can achieve better discrimination with a combined scoring paradigm. Since the scoring functions are not strongly correlated, as shown in the \hyperref[corr]{Appendix}, we hypothesize that a combined method may boost performance. 

\begin{figure}[h!]
  \hspace*{1cm}\includegraphics[width=0.9\textwidth]{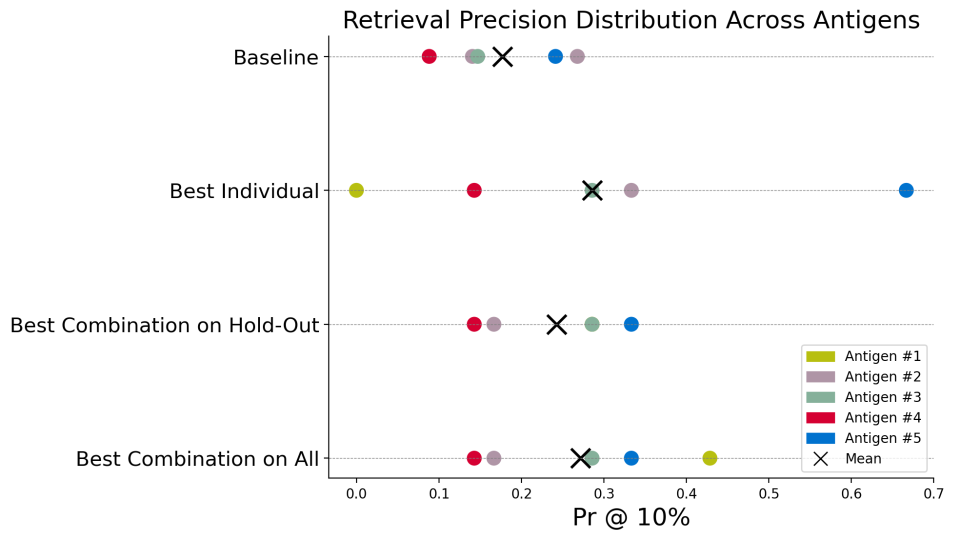}
  \caption{Distribution of Precision @ $10\%$ for ensemble scores and baselines across $5$ antigens.}
  \label{Figure5}
\end{figure}

For any linear combination of metrics, we fit coefficients on $4$ out of $5$ antigens, where our objective is mean AUROC. We do not consider combinations whose standard deviation exceeds $0.1$. To avoid leakage, we evaluate the best-performing ensemble score on the held-out antigen. We call this approach ``Best Combination on Hold-Out,'' and report retrieval precision on the task of filtering to $10\%$ of the original antibody design library's size in Figure \ref{Figure5}. Retrieval precision, or Precision @ $K$, measures the proportion of true binders in the top $K$ highest-scoring candidates. Retrieval precision and recall curves for the individual scoring metrics are given in the \hyperref[retrieval]{Appendix}. We also evaluate performance on an oracle approach where we find the best ensemble score on all $5$ antigens, which we refer to ``Best Combination on All.'' Our baseline is given by the binder percentage in each of the libraries, and performance of the best individual metric (antibody-aligned RMSD) is also shown for comparison.

We observe that ensembling does not improve performance. However, the standard deviation of the hold-out ensemble scores is $0.074$, compared to $0.22$ for antibody-aligned RMSD. This indicates that ensembling is more robust to antigen variation. Additionally, the hold-out ensemble scores are just $2.9$ percentage points below the oracle for retrieval precision, revealing that we cannot perform better on this classification task using these scoring criteria. This underscores the need for better prediction methods beyond those evaluated here. We find that in almost all folds, interface proximity has the highest weight. The specific weights are given in the \hyperref[weights]{Appendix}. This reveals that flexible docking information is critical for this task, and better antibody-antigen docking models can push the needle.

\section{Conclusions}

This study provides experimental insight to show that there is no one-size-fits-all approach for antibody assessment, and current open-source tools are insufficient to reliably filter diverse antibody libraries. We learn that antibody-aligned RMSD outperforms other individual scoring paradigms. However, sequence and undocked structure-only scoring paradigms struggle to handle antigen variation, and metrics that use antigen information are more robust. We find that linear combinations of the individual scoring paradigms exhibit lower variance in performance but leave much to be desired. Improved flexible docking models and further experimental benchmark efforts are critical to identify promising antibody candidates \textit{in silico} and accelerate their clinical development.

\section*{Acknowledgments}
The authors wish to thank Kieran Didi, Amaro Taylor-Weiner, Christine Lemke, Byron Olsen, and Kristin Iannacone for critical review of this manuscript.

\section*{Competing interest statement}
The authors are current or former employees, contractors, interns, or executives of Absci Corporation and may hold shares in Absci Corporation.

\newpage

{\small
\bibliography{references}
}

\newpage

\appendix
\section{Dataset details}

Table \ref{table:1} shows the number of screened antibodies and the percentage of binding antibodies for each library.

\begin{table}[h]
\begin{center}
\renewcommand{\arraystretch}{1.2} 
\begin{tabular}{ |c|c|c|c|c|c| } 
\hline
& \textbf{Antigen \#1} &\textbf{ Antigen \#2} & \textbf{Antigen \#3} & \textbf{Antigen \#4} & \textbf{Antigen \#5}  \\
\hline
\textbf{\# of Sequences} & $64$ & $56$ & $68$ & $68$ & $58$ \\
\textbf{\# of Binders} & $9$ & $15$ & $10$ & $6$ & $14$ \\
\textbf{\% Binders} & $14.1$ & $26.8$ & $14.7$ & $8.8$ & $24.1$ \\
\hline 
\end{tabular}
\vspace{0.8em}
\caption{Five libraries, each designed and screened against one of five antigens using surface plasmon resonance (SPR), are used as experimental data in this study. Each library consists of between 56 and 68 unique sequences with between 8.8\% and 24.1\% of the sequences binding to their target antigen.}
\label{table:1}
\end{center}
\end{table}

\section{dyMEAN training}
\label{dymean}
We make the following modifications to the training procedure of dyMEAN. While we train on the Structural Antibody Database \cite{dunbar2014sabdab} as in the original dyMEAN paper, we split SAbDab into training and validation sets according to clusters formed by complexes sharing $\geq 40\%$ antigen sequence similarity. The clusters containing any of the $5$ evaluation antigens were dropped from both training and validation. Additionally, we use OpenMM \cite{eastman2017openmm} to relax the bound antigen structures before model input to ensure that we are not leaking information about docked position. Finally, we provide the model with relaxed ABB2-predicted antibody structures to focus on learning flexible docking rather than folding.

\section{Docked complex scoring paradigms}
\label{docked-scoring}
We benchmark three scoring functions computed on docked complexes predicted by dyMEAN. The first is an interface proximity score which gives the average distance between the residues on the antibody interface and their nearest corresponding antigen residues. Here, we define interface as the contiguous region between the first residue of HCDR1 to the last residue of HCDR3, including HFWR2 and HFWR3. Residue-residue distances are computed as minimum distance between the side chain atoms of both residues. We also compute a contact proximity score which computes the same score on just the antibody contact residues which are defined as having at least one side chain atom within $5$ Angstroms of an antigen atom.

Lastly, we compute an interface interactions score which evaluates the types of interactions at the antibody-antigen interface. We begin by identifying all antibody-antigen residue pairs where an antibody residue atom is within at most 3 Angstroms of an antigen residue atom. We then classify each residue-residue pair, awarding higher scores to polar-polar interactions, nonpolar-nonpolar Van der Waals interactions, positive-negative ionic interactions, and aromatic stacking interactions.

\section{Correlation analysis}
\label{corr}
\begin{figure}[h!]
  \centering
  \includegraphics[width=0.9\textwidth]{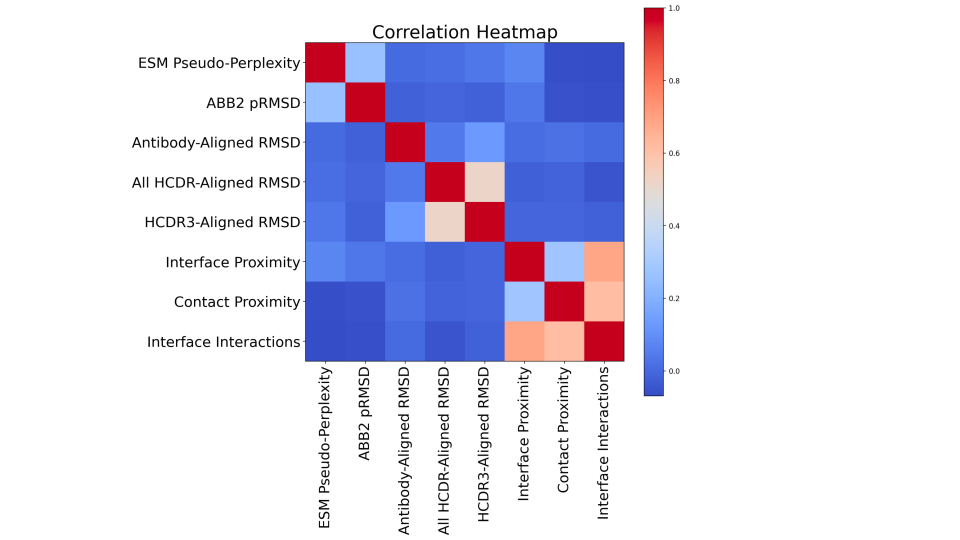}
  \caption{Correlations between the $8$ scoring paradigms, averaged across antigens, show that each metric is capturing fairly different signal. Combining metrics could yield favorable performance.}
  \label{Figure4}
\end{figure}

 Figure \ref{Figure4} shows correlations between the $8$ scoring paradigms. Correlation coefficients were computed on all pairs of each antigen's set of scores, and these coefficients were averaged across all $5$ targets. As expected, the RMSD-based undocked structure metrics exhibit some correlation, and the metrics computed on flexibly docked complexes are also correlated. With the exception of ESM pseudo-perplexity and ABB2 pRMSD, which are weakly correlated, we do not observe much correlation across the sequence, undocked structure, and docked complex metric groups. Therefore, we hypothesize that combining individual metrics may reveal new information useful for binder classification.

\section{Retrieval precision and recall}
\label{retrieval}
\begin{figure}[h!]
  \centering
  \includegraphics[width=0.9\textwidth]{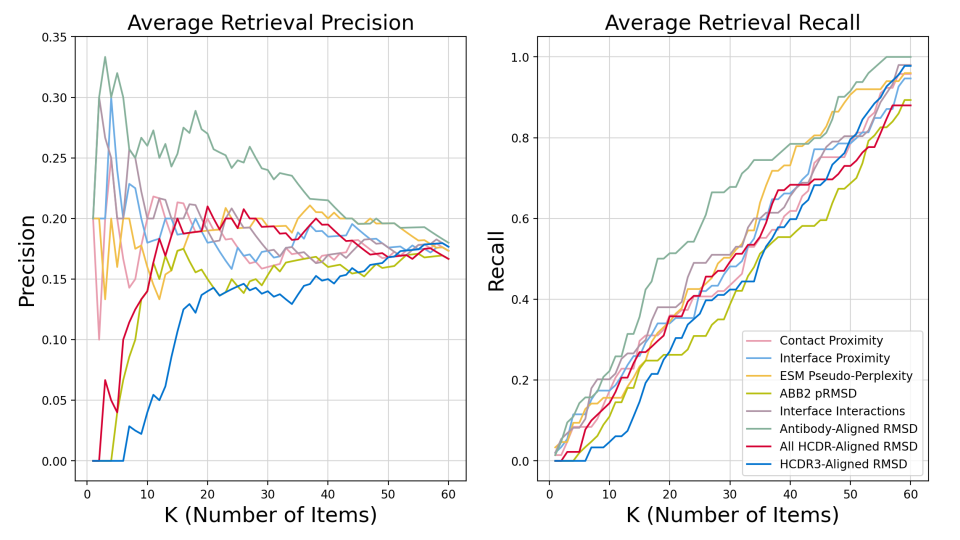}
  \caption{The retrieval precision and recall curves, averaged across antigens, show how effectively the different scoring paradigms can select $K$ designs. At a threshold of $10\%$, which is approximately $K = 6$, antibody-aligned RMSD displays the best retrieval precision and recall.}
  \label{Figure3}
\end{figure}

Retrieval precision and recall metrics are particularly relevant because  we can measure the number of true binders amongst the $K$ we may select to test in the lab, while ensuring that a high proportion of binders are not missed. Precision @ $K$ measures the proportion of true binders in the top $K$ highest-scoring candidates, and Recall @ $K$ measures the proportion of relevant items in the top $K$. We evaluate retrieval precision and recall values for each scoring function averaged across the $5$ antigens, as shown in Figure \ref{Figure3}. We focus on the task of filtering to $10\%$ of the original antibody design library's size. We observe that antibody-aligned RMSD has the highest Precision @ $10\%$ with a value of $0.26$, averaged across antigens, compared to a baseline mean binder rate of $0.13$. Antibody-aligned RMSD also has the highest retrieval recall of $0.16$ at this threshold.

\section{Combined scoring function weights}
\label{weights}
In Table \ref{Table3}, we report the coefficients found when each group of $4$ antigens was used for optimization.

\begin{table}[htbp]
\begin{center}
\renewcommand{\arraystretch}{1.2} 
{\fontsize{8}{10}\selectfont
\begin{tabular}{ |c|c|c|c| } 
\hline
\textbf{Held-out Antigen}& \textbf{Metric 1} & \textbf{Metric 2} & \textbf{Metric 3} \\
\hline
\textbf{Antigen \#1} & Interface Proximity, $0.7$ & Contact Proximity, $0.2$ & Interface Interactions, $0.1$\\
\textbf{Antigen \#2} & Antibody-Aligned RMSD, $0.5$ & Interface Interactions, $0.4$ & ESM Pseudo-Perplexity, $0.1$ \\
\textbf{Antigen \#3} & Interface Proximity, $0.7$ & Contact Proximity, $0.2$ & Interface Interactions, $0.1$\\
\textbf{Antigen \#4} & Interface Proximity, $0.7$ & Contact Proximity, $0.2$ & Interface Interactions, $0.1$\\
\textbf{Antigen \#5} & Interface Proximity, $0.7$ & ESM Pseudo-Perplexity, $0.2$ & Contact Proximity, $0.1$\\
\textbf{Oracle} & Interface Proximity, $0.6$ & Contact Proximity, $0.3$ & HCDR3-Aligned RMSD, $0.1$\\
\hline 
\end{tabular}
}
\vspace{0.8em}
\caption{Best weighted ensemble scores for each of the $5$ folds, as measured by mean AUROC.}
\label{Table3}
\end{center}
\end{table}

\end{document}